# Sodium-Ion-Conducting Polymer Nanocomposite Electrolyte of TiO$_2$/PEO/PAN Complexed with NaPF$_6$

Chandni Bhatt, Ram Swaroop, Parul Kumar Sharma and A. L. Sharma*

*Centre for Physical Sciences, Central University of Punjab, Mansa Road Bathinda-151001, Punjab, India*

*alsharmaiitkgp@gmail.com

**Abstract**. A free standing transparent film of solid state polymer electrolyte based on PEO/PAN + NaPF$_6$ with different compositions of nano sized TiO$_2$ in weight percent (x = 0, 1, 2, 5, 10, 15, 20 ) is synthesized by using standard solution cast technique. The homogeneous surface of above polymer composition is examined by FESEM. The microscopic interaction among polymer, salt and nanoceramic filler has been analyzed by Fourier Transformed Infra-Red (FTIR) spectroscopy. The reduction of ion pair formation in polymeric separator is clearly observed on addition of nanofiller in the polymer salt complex film. Electrical conductivity has been recorded of the prepared polymeric separator which is of the order of ~$10^{-3}$ Scm$^{-1}$ after addition of nanofiller (15% wt/wt) which support the FTIR results. Electrochemical potential window has been observed of the order of ~6V by the cyclic voltammetry results. The observed data of the prepared separator are at par with the desirable value for device applications.

## INTRODUCTION

Sodium is light alkali metal with single electron in outermost shell. This metal exhibit unexpected complexity properties [1]. Sodium is more environmentally friendly than Li, less expensive, more abundant, easily distributed and easier to extract than lithium [2]. Sodium is fourth most abundant element in earth crust and easy to recycle, used for sodium ion batteries. Sodium batteries provide low cost energy storage device, they operate at ambient temperature [3]. Polymer electrolyte act as electrolyte cum separator material in batteries, their main purpose is to separate anode and cathode material in batteries and act as ions transport medium for the conduction of ions. In most system conductivity much lower than the desirable values for the device application under surrounding conditions. The factor responsible for that are: low ambient ionic conduction, concentration polarization, poor stability (thermal, mechanical and electrochemical) [4]. To increase ionic conductivity various method employed such as, addition of nano-sized inorganic ceramic fillers particles such as Al$_2$O$_3$, SiO$_2$, TiO$_2$, ZrO$_2$ etc. in host polymer and other with the addition of low molecular weight plasticizers such as ethylene carbonate (EC), propylene carbonate (PC), dimethyl carbonate (DMC), polyethyleneglycol (PEG) [5]. In this work, investigation of solid state polymer electrolyte cum separator is fabricated by taking various composition of TiO$_2$ nano-filler adding in PEO/PAN polymer blend complexed with anatomic ratio of (O/Na = 20) NaPF$_6$ salt.

## EXPERIMENTAL WORK

In experiment the pure polymer blend of composition (PEO + PAN) films complexed with NaPF$_6$ salt prepared by solution cast technique. The weight percent ratio such as 50 weight percent of PEO, PAN, and salt of O/Na ratio is 20 mixed by a solution casting technique using DMF as solvent. Whereas in another samples the same polymer blend doped with different composition of nano-filler (TiO$_2$) was prepared in different proportion such as 1, 2, 7, 10, 15 and 20 weight percent. The solution of above composition mixed for several hours by means of stirrer, until the solution mixed properly. The obtained polymer blend nanocomposites were cast onto polypropylene dishes and left for evaporation at room temperature. The final product of polymer blend nanocomposites was dried under vacuum at $10^{-3}$ mbar pressure for 2-3 days. Finally we got free standing film of pure and doped polymer electrolyte.

Impedance spectroscopy was carried out for conductivity measurement with CH Instruments electrochemical workstation over the frequency range 1MHz, the sample sandwiched between stainless steel electrodes. FTIR analysis of the sample was done to detect the presence of various functional group. Surface morphology of the sample was observed by scanning electron microscopy. Working voltage range evaluated by cyclic voltammetry.

## RESULT AND DISCUSSION

### Complex Impedance Spectroscopy

Figure 1 represents complex impedance plot of prepared thin polymer nanocomposite films comprising of film $(PEO/PAN)_{20}NaPF_6+x$wt.%$TiO_2$. From fig1, it is observed that the higher frequency semicircle and low frequency spike present in both the sample. The higher frequency semicircle represents bulk contribution of the materials and spike due to the capacitance developed at electrode electrolyte interface. A nonlinear least square fitting of impedance response of all samples agrees well with electrical equivalent circuit model comprising a series combination of constant phase element (CPE) with another constant phase element (CPE) and resistance ($R_b$), which is parallel with one another, further they are in series combination with another resistance ($R_1$).

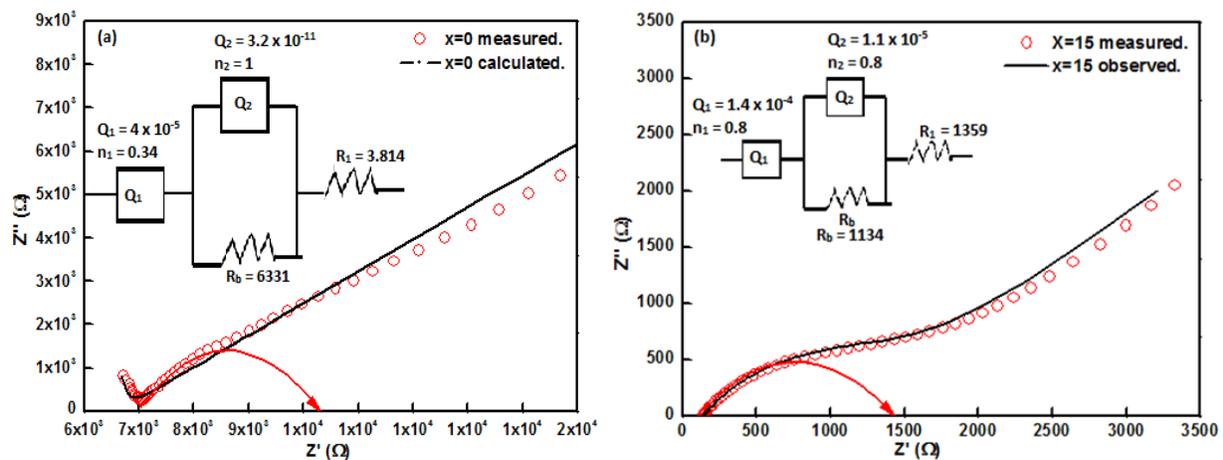

**FIGURE. 1**. Nyquist plot of thin PNC films comprising of $(PEO/PAN)_{20}NaPF_6+x$wt.%$TiO_2$ (a) x=0 (b) x=15

The conductivity ($\sigma$) of the polymer electrolyte was calculated from the following formula. $\sigma = L/R_bA$, where L is thickness of polymer electrolyte film (cm), and A is the area of crossection of electrode (cm$^2$) [6]. The electrical conductivities values for blend polymer and $TiO_2$ doped blend polymer films are recorded in the table 1.

Table 1. Conductivities values for polymer blend PEO/PAN complexed with $NaPF_6$ with different concentration of nano-filler $TiO_2$ (0, 1, 2, 5, 10, 15, 20) for sodium ion batteries.

Table1 clearly indicates that electrical conductivity has improved by an order value on after addition of 15wt.% $TiO_2$ in pure blend polymeric films.

| Polymer Electrolyte | Nano-filler $TiO_2$ (in wt. percent) | Bulk Resistance ($R_b$) In (Ω) | Conductivity (S/cm) |
|---|---|---|---|
| (PEO/PAN + $NaPF_6$) | 0 | 6331 | $3.2 \times 10^{-5}$ |
| (PEO/PAN + $NaPF_6$) | 15 | 1134 | $1.8 \times 10^{-4}$ |

**Electrochemical Analysis**

For electrochemical devices such as batteries, capacitors and electrochromic devices the electrochemical stability window curve is prime requirement. The cyclic voltammetry has been performed of the prepared cathode materials. This process was observed in the potential range of ~6V at current density of 0.00004 to -0.00025 through blocking electrode of stainless steel. Electrochemical stability window curve is obtained in the range from about ~6V, which is an acceptable voltage range for device application [7].

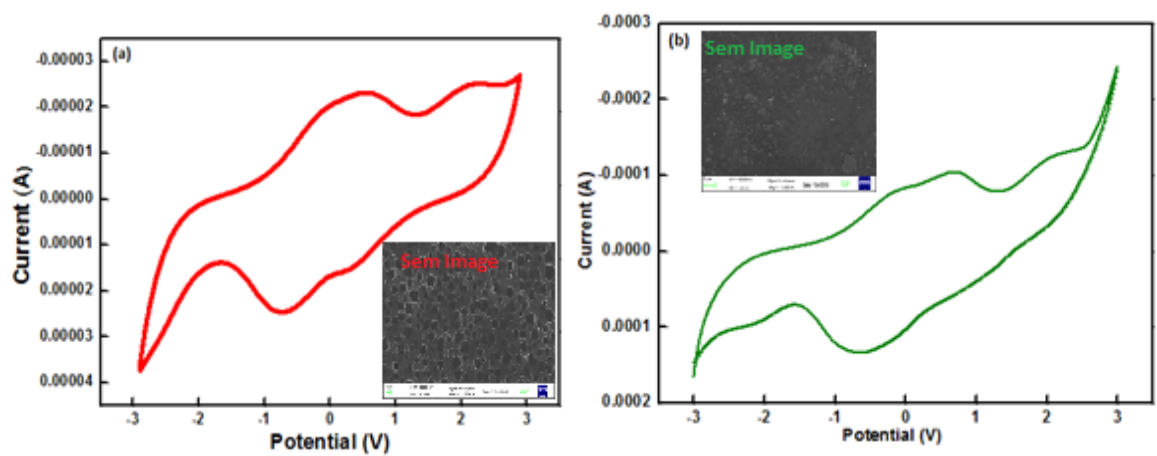

**FIGURE. 2.** Cyclic voltammeter images of (a) pure (PEO/PAN+$NaPF_6$) (b) doped15wt. % $TiO_2$ + ((PEO/PAN+$NaPF_6$) polymer electrolyte.

**Fourier Transform Infrared (FTIR) analysis**

Fourier Transfer infrared microscopy analysis of the sample was done to detect the presence of various functional groups. FTIR spectrum was observed in the range of 600-4000 $cm^{-1}$ as shown in fig.3. Also shift in peak position with addition of salt was clearly seen infig.3 (a). The peak in the wavenumber range 800 $cm^{-1}$ to 900 $cm^{-1}$ corresponds to C-O-C of PEO and salt ($PF_6^-$). The characteristic absorption peak observed in the spectral pattern at the wavenumbers ~690, ~840-920, ~1088, ~ 1247, ~1670, ~ 2242, ~2856 and 2932 $cm^{-1}$ are attributed to def(CH2), ($PF6-$), t(CH2)s, t(CH2)as, $\upsilon(C=O)$, $\upsilon(C\equiv N)$, (CH2) and (CH2) respectively. Two clear $CH_2$ vibrational modes appear due to PEO near 1458 $cm^{-1}$ which, correspond to asymmetric $CH_2$ bending (δ(CH2)a) and near 1356 $cm^{-1}$ which, corresponds to symmetric $CH_2$ wagging and some C- C stretching (w(CH2)s + ν (C—C)). The FTIR peaks at 1958 $cm^{-1}$ are related to asymmetric stretching of $CH_2$ in PEO. The C≡N stretching band in the IR spectrum appearing at 2245 $cm^{-1}$is the most characteristic feature of nitrile group in pure PAN. The vibrational peak at 2935 $cm^{-1}$, 2938 cm−1, 2921 cm−1 and 2923 $cm_{-1}$ are assigned to asymmetric stretching of $CH_2$. As $NaPF6$ salt concentration increases, there is shift in peak position of pure PEO/PAN+ $NaPF6$ sample related to other $TiO_2$ doped sample as shown in fig. (b), this is due to the additional amount of Na+ in the complexes. Pure PEO shows a large, broad band of $CH_2$ stretching between 2950 and 2840 cm-1. The FTIR analysis gives us the information of role of salt in the polymer and salt interaction in the polymer film.

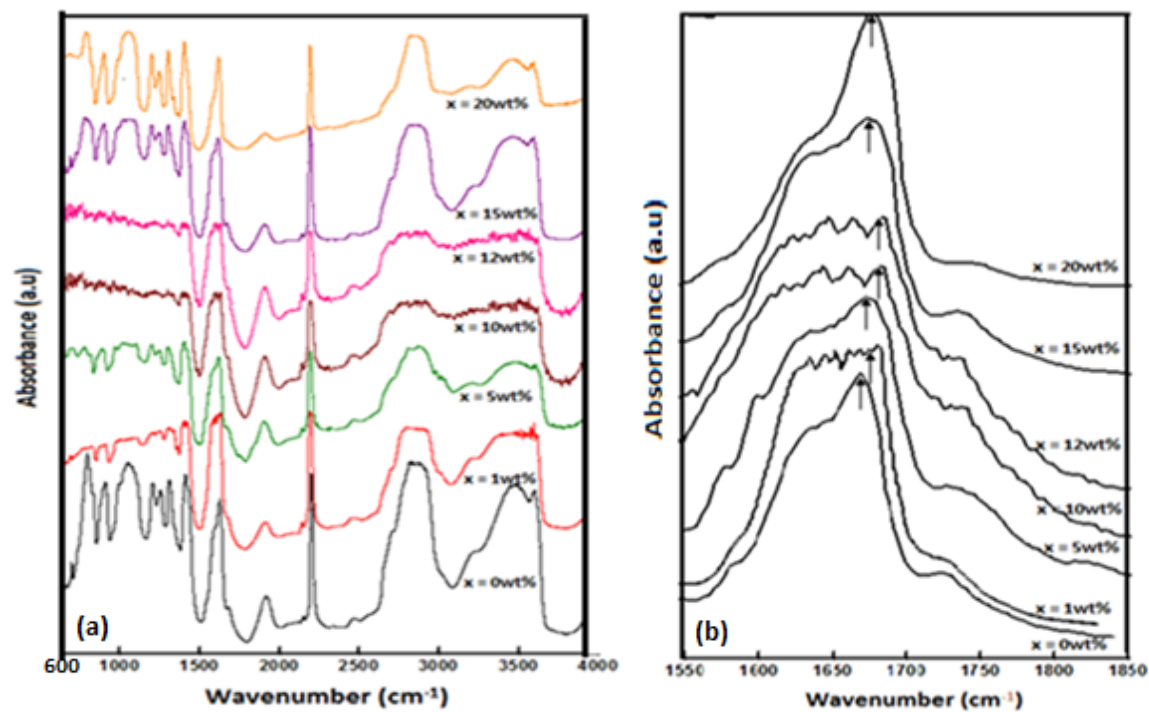

**Figure. 3.** FTIR spectrum for (a) different composition for x wt. % of $TiO_2$ (b) Variation of PEO peak at 1600 cm$^{-1}$.

## CONCLUSION

Free standing film was prepared by using PEO/PAN as polymer blend complexed with $NaPF_6$ salt with different concentration of $TiO_2$ (0, 1, 2, 5, 10, 15, 20) as nano filler. From impedance spectroscopy the value for maximum conductivities is 1.8 X 10$^{-4}$ Scm$^{-1}$ for 15wt. % of $TiO_2$ concentration which further decrease with the more addition of nano-filler. Electrochemical stability for our film is obtained in the potential range from ~6V which is good for device application. FTIR spectroscopy confirms the salt, polymer, and nano-filler interaction.

## ACKNOWLEDGEMENT

Author is thankful to Central University of Punjab for providing material, instrumental and Laboratory facility.